# An Algorithm to Speed up the Spatial Power Profile Calculation in Backward Raman Amplified Systems

Jad SARKIS, Yanchao JIANG, Pierluigi POGGIOLINI

Optcom, Politecnico di Torino, Italy

**Abstract:** As data transmission demands grow, long-haul optical transmission links face increasing pressure to increase their throughput. Expanding usable bandwidth through Ultra-Wide Band (UWB) systems has become the primary strategy for increasing transmission capacity. However, UWB systems present challenges, such as the reliance on backward Raman amplification and the complications posed by inter-channel stimulated Raman scattering (ISRS), which causes uneven signal propagation across bands. To address these issues, accurate and efficient physical models are required for real-time optimization, which rely on the knowledge of the power profile. This paper develops a novel, more efficient method for computing the power profile of signals and pumps, utilizing the integral form of the equations with matrix-based approximations. The algorithm achieves up to a thirty-fold average speed increase over conventional approaches while maintaining an error margin under 0.05 dBm. These results represent a significant step forward towards reducing optimization times and enabling more extensive studies in ultra-wide band long haul optical transmission, further facilitating research and commercialization of UWB systems, in an effort to address the growing demand for higher throughput.

## 1. Introduction

The exponential growth in data transmission demands is driven by different factors, such as the increasing numbers of users and the development of more demanding applications. This growth has placed significant pressure on the long-haul optical transmission links to increase their provided throughput. To accomplish this increase, researchers are focusing on two possible approaches: Increasing the bandwidth efficiency by transmitting more bits per unit of frequency or increasing the exploitable bandwidth. The first method requires more complex modulation formats thus faces distance limitations in long-haul transmission. As a result, bandwidth expansion is becoming the primary solution to meet rising transmission needs.

The objective of this paper is to investigate Ultra-Wide Band (UWB) systems, which span across multiple optical frequency bands, including the C, L, S and E bands. These systems can provide up to 30 THz of exploitable bandwidth, significantly exceeding the 5 THz available in the C band alone, traditionally used in optical transmission. While UWB systems offer the potential for substantial throughput gains, they also introduce new challenges. One of these is Inter-channel Stimulated Raman Scattering (ISRS), a nonlinear effect that becomes significant between widely spaced channels, causing power transfer from higher-frequency channels to lower-frequency ones. Furthermore, as demonstrated in [1], backward Raman amplification becomes necessary to achieve higher throughput and maintain flat GSNR curves across the entire usable spectrum. Consequently, many input parameters such as individual channel and pump input power and pump frequency need to be optimized to mitigate the effects of ISRS and maximize link throughput. For this reason, a physical model able to predict the achievable general signal to noise ratio (GSNR) on the link according to the input parameters is a must for the optimal operation of such systems. The most prominent physical models already in use in the industry are Closed Form Models (CFM) of the GN and EGN models developed by Politecinco di Torino in collaboration with CISCO in [2] and [3], the last of which has been extensively validated experimentally in [4].

However, one part of the CFM remains dependent on numerical calculations: The computation of the power profile. That is determining the power of each channel and pump at every point in the fibre, while taking ISRS and backward Raman amplification into account. Moreover, in backward Raman amplified systems, power profile calculations become a double-boundary value problem and the solving of the coupled Raman differential equations using standard numerical methods becomes inefficient. Because of this, the computation of the power profile using standard numerical methods consists of the bottleneck of the CFM, consuming a staggering 95% of the model's computation time.

The main objective of this paper is to develop a more efficient algorithm to compute the power profile, that is both fast and reliable compared to the conventional method. The successful development of such an algorithm will significantly reduce the computation time required to assess the achievable GSNR on any given link. This, in turn, will enable more efficient and large-scale optimization of UWB systems, facilitating future research of such systems. In practical terms, this advancement would allow network operators to manage and optimize multi-band systems in real-time, ensuring robust performance even under dynamic network conditions. By enhancing the speed of power profile calculations, this algorithm will lay the groundwork for more widespread adoption of UWB technologies, ultimately helping to meet the growing demand for high-throughput long-distance optical communication systems. In this paper, the proposed approach computing the power profile using the integral form of the Raman equations will be described, after which the algorithm building on this approach will be presented along with the respective performance and efficiency results.

## 2. The proposed algorithm

### 2.1 Key concept

We start from the coupled differential Raman equations in two directions for arbitrary Raman amplifications, and the ISRS effect. The power evolution within the $i$-th channel is modeled as,

$$\pm \frac{dP_i(z)}{dz} = -\alpha_{0,i} P_i(z) + \sum_{j=1}^{M+N} g_{ij} P_j(z) P_i(z) \tag{1}$$

The symbols '±' represent the two directions of propagation, where '+' denotes forward-propagating signals or pumps, and '-' denotes backward-propagating pumps. There are $M$ signal channels and $N$ pump channels. The $i$-th channel is characterized by its center frequency, denoted as $f_i$, and $f_i < f_j, 1 \leq i < j \leq M + N$.

These equations state that the power variation of a channel $i$ in an increment of the fibre length $dz$, is determined by the proportion of power lost to attenuation and that of power lost or gained due to ISRS or Raman pumps amplification. The upper and lower sign indicate a co-propagating wave (signals or co-propagating pumps) and a counter-propagating wave (backward pumps) respectively.

The main problem in numerically resolving this system of differential equations lies in the fact that the system is a double-boundary value problem where boundary conditions are imposed at both ends of the fiber. This requires multiple iterations along the whole fiber length to re-conciliate both boundary conditions. Moreover, due to the physical properties of the system, the variation of channel power can change drastically in different regions of the fiber. Instead, another approach was proposed by [5], by moving to the integral form of the equations, seeking solutions along the iteration axis rather than the fiber propagation axis.

By dividing both parts of Eq. (1) by $P_i(z)$ and integrating, the following integral form results:

$$P_i(z) = P_i(0) \cdot \exp\left(\mp \alpha_{0,i} z \pm \sum_{j=1}^{M+N} g_{ij} \int_0^z P_j(\psi) d\psi\right) \tag{2}$$

In this form, the power profile of channel $i$ at any point $z$ of the fiber, $P_i(z)$, is explicitly expressed as a function of $z$. The difference is that the coupled differential equations are no longer solved directly, which involved multiple variables affecting each other's rate of change. Rather the solution is obtained by iterating through a formula of $P_i(z)$ where each value is updated based on the values obtained from previous steps.

A vector $Z$ is created to represent all the steps inside the fiber, with a given step size $\Delta Z$,

$$Z = [0, \Delta Z, 2\Delta Z, \ldots, L - \Delta Z, L] \tag{3}$$

where $L$ is the total length of the fiber. The number of steps is defined as $N_z = \lceil L/\Delta Z \rceil$ and the total number of channels including signals and pumps is defined as $N_{ch} = M + N$.

Then, the power profile of all channels can be expressed as,

$$P = \begin{pmatrix} P_1(0) & P_1(\Delta z) & \cdots & P_1(L) \\ P_2(0) & P_2(\Delta z) & \cdots & P_2(L) \\ \vdots & \vdots & \ddots & \vdots \\ P_{N_{ch}}(0) & P_{N_{ch}}(\Delta z) & \cdots & P_{N_{ch}}(L) \end{pmatrix} \tag{4}$$

Furthermore, the integral present in the equation must be numerically solved. This is accomplished using the trapezoidal rule, which approximates the integral by dividing the area under the curve into trapezoids. Mathematically, this approximation can be written for an interval $[Z_i, Z_{i+1}]$ as:

$$\int_{Z_i}^{Z_{i+1}} P_j(\psi) d\psi \approx \Delta Z \frac{\left(P_j(Z_i) + P_j(Z_{i+1})\right)}{2} \tag{5}$$

Repeating it for each interval along the entire link, we can get,

$$\begin{aligned} \int_0^z P_j(\psi) d\psi &\approx \Delta Z \frac{\left(P_j(0) + P_j(\Delta Z)\right)}{2} + \Delta Z \frac{\left(P_j(\Delta Z) + P_j(2\Delta Z)\right)}{2} + \ldots + \Delta Z \frac{\left(P_j(z - \Delta Z) + P_j(z)\right)}{2} \\ &= \Delta Z \left(\frac{1}{2} P_j(0) + P_j(\Delta Z) + P_j(2\Delta Z) + \ldots + P_j(z - \Delta Z) + \frac{1}{2} P_j(z)\right) \end{aligned} \tag{6}$$

Finally, the vector equation that computes the power profile vector, containing the powers of a given channel $i$ sampled at equal steps $\Delta Z$ of the fiber becomes,

$$\int_0^z P_j(\psi) d\psi \approx \Delta Z \cdot P_j \cdot T_{trig} \tag{7}$$

$$T_{trig} = \begin{bmatrix} 0 & 1/2 & 1/2 & 1/2 & \ldots & 1/2 \\ 0 & 1/2 & 1 & 1 & \ldots & 1 \\ 0 & 0 & 1/2 & 1 & \ldots & 1 \\ 0 & 0 & 0 & 1/2 & \ldots & 1 \\ \vdots & \vdots & \vdots & \vdots & \ddots & \vdots \\ 0 & 0 & 0 & 0 & \ldots & 1/2 \end{bmatrix}$$

The above equation can be extended to compute the whole power profile of all channels in one shot,

$$P = P(:,0) \cdot \exp\left(-Direction \cdot \alpha_0 \cdot Zk + Direction \cdot G \times P \times T_{trig} \cdot \Delta Z\right) \tag{8}$$

Equation (8) serves as the foundation for the efficient power profile computation method presented in this paper. Starting with an initial guess of the power profile and applying the boundary conditions, this equation allows for the propagation of the profile according to the Raman differential equations, providing a numerical solution for the system in vector form. However, due to the numerical nature of the method, the accuracy of the output depends heavily on how close the initial guess is to the actual solution.

To determine when an algorithm has converged, it is essential to define what constitutes a "satisfactory solution." A correct solution must first remain stable when propagated through Eq. (8). Additionally, if the profile obtained after propagation matches the boundary conditions within a certain error tolerance, it is considered valid. Since the initial guess is rarely accurate enough to produce a satisfactory solution on the first try, an iterative approach is necessary. With each step, the obtained profile is adjusted slightly for the next iteration, and the boundary conditions are monitored to ensure convergence towards the correct solution. Three methods that implement this process in different ways, namely dynamic pump allocation method, progressive signal injection method and the proposed method will be described in the following sections.

## 2.2 The proposed algorithm

The first two methods developed initially, dynamic pump calibration (DPC) and progressive signal injection (PSI) were based on the integral form of the equations but followed different strategies to ensure convergence to a solution. Broadly, DPC works by iteratively correcting the pump powers proportionally to their algebraic distance from their boundary conditions. While PSI starts from heavily scaled down signals and increases their power by a small step in each iteration until their initial conditions are reached. In both methods and in each iteration, the corrected profiles are propagated in a forward fashion using the vector equation.

While both methods performed well, each one faced specific limitations. DPC struggled with highly dynamic parameters that were difficult to adapt to every scenario, whereas PSI encountered accuracy issues, lacking control over precision metrics. The core contribution of this paper is the development of a hybrid method that combined these two approaches, transitioning from PSI to DPC in a serial manner, where PSI's final solution is passed to DPC as an initial guess. This integration allowed the hybrid method to retain the strengths of both while overcoming their respective shortcomings.

Initially, the final algorithm is reported in Table 2.1 for the convenience of the reader, so one can follow its different steps right from the start. The algorithm begins by initializing the power profile for signals going from their initial condition and propagating through the fiber subject only to attenuation. While the power profile of the pumps starts from the other end of the fiber at the boundary value and travels the fiber not only subject to attenuation but also to the Raman effect between the pumps themselves, this is achieved by solving the Raman differential equations for the pumps only, forming an initial value problem, using an ordinary differential equation solver. To complete the initial power profile, both the signal and pump power profiles are scaled-down, using defined scaling factors, $factor_{signal}$ and $factor_{pump}$ for the signals and pumps as well as step sizes $step_{dBm, signal}$ and $step_{dBm, pump}$ for incremental power increases later on in the algorithm.

This is done for different reasons in each case. For the pumps, it can be shown that the presence of high power in the system can lead to divergence after a small number of iterations. Thus, the pumps, typically containing the most amount of power in the system, are scaled down at the beginning to avoid early divergence. For the signals, the scaling down comes from the idea that instability in the algorithm seems to be due to the interaction between signals and pumps; the more power is exchanged between the two, the more the equations are unstable. For this reason, the method starts under the assumption of un-depleted pumps, meaning pumps that have not transferred power to the signals. Since the power transfer between two channels is proportional to the power of both channels, this is accomplished by injecting signals having negligible power. This is achieved by scaling down the signals by $factor_{signal}$.

The method then employs a two-stage loop structure to solve the problem: in the first stage, the method proceeds by moving the signal power up by steps of $step_{dBm, signal}$, a parameter of the method. In each iteration, the signal power is moved up, and the guess is propagated through the vector equation. Moreover, because some interaction is happening between the pumps and the signals, the pump power profiles do get affected by the propagation. In each iteration this should be corrected by scaling the pumps, so they return to their scaled boundary conditions exactly, like the following:

$$P(N_{ch_{signal}}+i,:) = P(N_{ch_{signal}}+i,:) \cdot \frac{P_{pump,in,scaled}(i)}{P(N_{ch_{signal}}+i,L)} \qquad (9)$$

It is noted that the boundary condition, $P_{pump,in}$ must also be scaled by $factor_{pump}$, resulting in the scaled boundary condition $P_{pump,in,scaled}$ used in the equation above for pump correction. This is done because the pumps themselves were scaled down in the first place.

The algorithm keeps going until the signal power rejoins the initial condition exactly. By increasing signal power slowly and correcting the pump profiles each time, the system is moving from one converged solution of the system to the other, allowing it to finally reach a correct solution with the given boundary conditions.

Once the signals have been fully scaled up, the pumps are then incrementally restored to their true boundary condition. Thus, both the pumps and their current boundary conditions $P_{pump,in,scaled}$ are scaled back in a step-by-step manner over several iterations using a parameter $step_{dBm,pump}$. In each iteration, the profile is propagated through the vector equation after the up scaling. Moreover, in each iteration, the pumps are corrected to align with the scaled boundary condition at that stage, $P_{pump,in,scaled}$ similarly to the above equation.

**Table 2.1** the proposed algorithm

```
Algorithm
1: Initialize: Scale down signal profile by factor_signal. Set step sizes step_{dBm,signal}
   and step_{dBm,pump}, and tolerance tol.
2: Scale down pump profile and pump boundary condition (P_{pump,in}) by factor_pump
3: P_{pump,in,scaled} = P_{pump,in} / factor_pump
4: Solve the initial pump-only power profile using ode45.
5: Initialize flag signal_scaled_up to 0.
6: while true do
7:     Correct pump profile:
8:     for each pump i do
9:         Scale pump profile to match scaled down boundary condition:
10:        P(N_{ch_signal}+i,:) = P(N_{ch_signal}+i,:) · P_{pump,in,scaled}(i)/P(N_{ch_signal}+i,L)
11:    end for
12:    if signal_scaled_up == 1 then
13:        if Any(|P_pump(L) - P_{pump,in}|) > tol then
14:            Increase pump power by step_{dBm,pump}
15:            Increase boundary condition P_{pump,in,scaled} by step_{dBm,pump}
16:        end if
17:        if All(|P_pump(L) - P_{pump,in}|) < tol then
18:            Propagate using the vector equation:
19:            P = P(:,0)·exp(-Direction · α_0 · Zk + Direction · G × P × T_{trig} · ΔZ)
20:            Feed PSI's guess to DPC method described in 1
21:            Break out of while loop when DPC converges
22:        end if
23:    else
24:        Increase signal power by step_{dBm} to form the new guess
25:    end if
26:    Propagate the guess using the vector equation:
27:    P = P(:,0) · exp(-Direction · α_0 · Zk + Direction · G × P × T_{trig} · ΔZ)
28:    if P_signal(0) has reached its target then
29:        signal_scaled_up = 1
30:    end if
31: end while
```

Once the pumps have been fully scaled back to their actual boundary condition $P_{pump,in}$, the entire profile is propagated through the vector equation, and the first stage of the method concludes. It is mentioned that, in the algorithm, *signal_scaled_up* is a flag that becomes true when the method has finished scaling up the signals signaling the initiation of the pump scaling up process.

The second stage of the method, shown in Table 2.2, begins when pumps are done being scaled up, in the part of the algorithm where the condition $All(|P_{pump}(L) - P_{pump,in}|) < tol$ is satisfied. The method takes the guess provided by the first stage and propagates it in the vector equations. It is seen that some pump values go above their respective boundary condition while others have decreased under it. As a result, the signal profile is moved away from the correct values. For this reason, this method focuses on getting better guesses for the pump profile to achieve an overall accurate power profile. Thus, the method fixes the signal values at the input of the fiber to the signal initial conditions and works on the pump profile by pushing each pump towards its boundary conditions.

Table 2.2: Dynamic Pump Calibration Method

```
Algorithm 1 Dynamic Pump Calibration Method
1:  Initialize: Set initial guess for power profile P and error tolerance tol
2:  while Any(|Pump_Error|) ≥ tol do
3:      Pump_Error(i) = P_pump,in(i) − P(N_ch_signal + i, L)
4:      for each pump i do
5:          if Pump_Error(i) > 0 then
6:              Correct pump under-calculation:
7:              P(N_ch_signal + i, :) = P(N_ch_signal + i, :) · (1 + CL) · Pump_Error(i)
8:          else
9:              Correct pump over-calculation:
10:             P(N_ch_signal + i, :) = P(N_ch_signal + i, :) · (1 + CH) · Pump_Error(i)
11:         end if
12:     end for
13:     Propagate new power profile:
14:     P = P(:, 0) · exp(−Direction · α_0 · Zk + Direction · G × P × T_trig · ΔZ)
15: end while
```

For this purpose, the method uses a metric that measures how far the pumps at the end of the fibre are from their respective boundary conditions called Pump_Error:

$$\text{Pump\_Error}(i) = P_{pump,in}(i) - P(N_{ch_{signal}} + i, L) \tag{11}$$

Where the first term is the boundary condition of pump $i$ in Watts and the second term is the calculated value at the output of the fiber of pump $i$ in Watts. It is noted that when the value of Pump_Error($i$) is positive, the pump is lower than it should be and has been under-calculated while in the opposite case it has been over-calculated. Consequently, based on the Pump_Error metric, the method will correct the power profile guess for the next iteration by adding to it a correction value obtained by multiplying the profile of each pump $i$ with a value proportional to Pump_Error($i$) in the following way:

$$P(N_{ch_{signal}} + i, :) = P(N_{ch_{signal}} + i, :) \cdot (1 + CL) \cdot \text{Pump\_Error}(i)$$
$$P(N_{ch_{signal}} + i, :) = P(N_{ch_{signal}} + i, :) \cdot (1 + CH) \cdot \text{Pump\_Error}(i) \tag{12}$$

The proportionality coefficient $k$ is called the correction factor. Moreover, a distinction is made between corrections of under-calculation and over-calculation, as the latter is more critical in divergent cases while the former is important in oscillation. This introduced two correction factors: Correction factor higher (CH) used in the case of over- calculation and Correction factor Lower (CL) used in the other case. An analysis of the correction factor values will be provided later. After correction is done, the profile is passed through the vector equation and the whole is repeated until convergence is achieved.

By doing this correction step, the method is changing the guess given to the vector equation to a better one, while keeping the boundary conditions the same, thus pushing the equation to give an estimation to the power profile slightly more accurate in each iteration.

The loop yields a converged result by fixing a value for the error tolerance, called $tol$, below which the pump power error at the end of the link is deemed satisfactory. After this condition is satisfied, the hybrid method converges to the desired power profile and the code breaks out of the while loop.

This section has described in detail the logic of the hybrid method by going over its two-stage functionality. The proposed method has many parameters that need to be fixed or made adaptive, in the following section, an analysis of all the parameters is provided, some of which are given a fixed value while two of them, namely *CL* and $factor_{pump}$ are made adaptive through two algorithms integrated to the algorithms presented above.

## 2.3 Parameter Automation

The main objective is to develop a method for efficient power profile calculation that can be inserted in the closed form GN model, to render it faster and enable it to be used in real-time applications and perform efficient complex optimization. These objectives require the developed method to solve different UWB systems in a vast space of signal and pump power. For this reason, the method's parameters should all be set to some value that works in most realistic cases or at least have an adaptive mechanism that adapts them to an acceptable value for each case scenario.

The Hybrid method described in the prior section has the following tunable parameters:

- CH=5: this parameter is not very dynamic. It is mostly useful to avoid divergence.

- CL

- $step_{dBm,signal}$ =2: this allows the method to be efficient by scaling up the signals quickly while maintaining a good accuracy.

- $step_{dBm,pump}$ = 0.5: this allows the method to be efficient by scaling up the pumps quickly while maintaining a good accuracy.

- $factor_{signal}$ = 4: Fast enough with no real effect on convergence or accuracy.

- $factor_{pump}$

- Stopping number = 3000. As most systems converge before this amount of iterations.

The given values may not be the optimal ones. Some optimizations could be done in future research to determine the optimal values, which would definitely have an impact on both efficiency and robustness. However, heavily tested values have been given that seem to be working well in most cases.

The 2 remaining parameters that on the other hand do not seem to have one specific value that can be reused every time are CL and $factor_{pump}$. This was seen by the fact that the highest percentage of power inside the system is injected through the pumps, meaning $factor_{pump}$ needs to be dependent on it.

Consequently, 2 simple algorithms are put into place to make these 2 parameters adaptive to the specific case the Hybrid method is solving. Each of which will be discussed below.

### 2.3.1 Pump Factor

when the method diverges because of too high pump power, it does so in a couple of iterations after which the loop is exited. Meaning that no significant loss of time happens if the system diverges and is rerun with a different value. The idea is to start the algorithm with no pump scale down (a value of 1), then if divergence happens, the system is rerun with a new value picked from a vector of Factors containing increasing values, for example Factors = [1 5 10 15].

Divergence is detected by checking the P matrix for NANs. This way, the factors in Factors are tried one by one. If the method could not converge for any of those it assumes that power is too high, and no convergence can be achieved. It breaks and signals what happened through the divergence flag. Otherwise, if no NANs are found in the converged P, *nanflag* is set to 0 and the loop is exited. The algorithm is shown in Table 2.3.

**Tabel 2.3**: automated pump factor adjustment algorithm

```
Algorithm    Automated Pump Factor Adjustment Algorithm
1:  Initialize: Define the Factors vector, k=1, factor_pump = Factors(k), nanflag=1
2:  while nanflag == 1 do
3:      Run algorithm in 4 with current factor_pump.
4:      Check for divergence:
5:      if any(isnan(P)) then
6:          nanflag = 0;                      ▷ No NANs, algorithm converged
7:      else
8:          k = k + 1;
9:          if k > length(factor_pumps) then
10:             divergence_flag = 1;    ▷ Algorithm diverged for all pump scales
11:             Break;
12:         else
13:             Update pump factor: factor_pump = factor_pumps(k);
14:         end if
15:     end if
16: end while
```

## 2.3.2 Pump Factor

The lower correction factor (CL) addresses the trade-off between oscillation and efficiency. If the value is too high, it can cause infinite oscillation with frequency proportional to the value of CL, while a very low value may result in slow convergence. The challenge lies in finding the right balance every time, as this trade-off shifts across different boundary conditions.

A good way to address this issue relies on early oscillation detection. The method can start with a big enough CL that works well for low pump powers, then a mechanism is put in place to detect oscillation as it is happening. The method can then lower the value of CL every time oscillation is detected.

To detect oscillation, a MATLAB function called "findpeaks" can be used on one of the Pump_Error curves, returning the values of the peaks and their position. Of course, not all peaks should be considered, but only those with significant magnitude. Oscillation is assumed when the number of peaks with a value bigger than $magnitude_{thresh}$ is bigger than a certain threshold $Peaks_{thresh}$. As a response, the CL value is reduced proportionally to the frequency of the oscillations as follows:

$$CL = \frac{CL}{c_0 \cdot length(peaks)} \qquad (13)$$

where $c_0$ is a proportionality coefficient that determines how sensitive to the oscillation frequency the CL reduction should be. It is initially set to 1 and left for future optimization.

Oscillation is checked every 100 iterations. At each check, MATLAB's "findpeaks" function is applied to one of the Pump_Error curves, examining the section from the last recorded CL adjustment (last_change) to the current iteration. Initially, last_change is set to 1 and updated whenever the CL is reduced. This approach ensures that oscillations that might emerge over a longer period can still be detected, while the peaks that triggered previous CL reductions are no longer considered in subsequent checks.

The algorithm works as in Table 2.4. The initial CL value should be determined, some test showed that a value of 0.1 seems to be working well for low pump power, thus the method starts with it and then reduces it as it sees fit.

**Tabel 2.4:** automated pump factor adjustment algorithm

```
Algorithm    Lower Correction Factor Adjustment Algorithm
 1: Initialize: Set initial correction factor CL, last_change = 1, oscillation check
    every 100 iterations, threshold values magnitude_thresh = 0.3 and Peaks_thresh = 2,
    and proportionality constant c_0 = 1.
 2: while algorithm is running do
 3:     if iter_number ∈ oscillation_check_values then
 4:         Extract pump error:
 5:         Pump_Error_1 = Pump_Error(last_change : iter_number, 1)
 6:         Find peaks in the Pump Error curve:
 7:         peaks = findpeaks(Pump_Error_1)
 8:         Filter insignificant peaks:
 9:         peaks = peaks(peaks > magnitude_thresh × max(Pump_Error_1))
10:         if length(peaks) > Peaks_thresh then
11:             Adjust correction factor:
12:             CL = CL / (c_0 × length(peaks))
13:             Update last change iteration:
14:             last_change = iter_number
15:         end if
16:     end if
17:     Continue with the rest of the algorithm.
18: end while
```

# 3. Performance Evaluation and Comparison

For the proper use of the method, it is important to understand its reliability across different scenarios, and its time-saving potential in comparison to traditional power profile computation method. The analysis includes extensive testing under extreme signal and pump power conditions and various ultra-wide band scenarios, along with a detailed comparison of speed, accuracy and convergence. Additionally, an overview of the average time gain achieved in an optimization scenario, relative to the conventional approach, will be presented.

## 3.1 Stress Test

A C+L system was used. This system has 38 channels in the C band and 38 channels in the L band, with frequencies spanning in the range [186;195] THz. The channel spacing is 125 GHz and a symbol rate of 100 GBaud is used. 5 counter-propagating pump channels are used in the frequencies: [210.56, 208.87, 206.72, 204.51, 200.55] THz, and the original powers: [ 360 320 200 130 180] mW. The power of each signal and pump channel at the respective ends of the fibre, forming the boundary conditions, are variable during the different tests to ensure the generalization of the method for different power scenarios.

For stress test, the method is used to solve C+L systems with adjustment factors going from 1 to 0.1 with steps of 0.1. It should be noted that the adjustment factor divides the original pump powers to scale them up or down, meaning that a low Adjustment factor indicates high pump powers. Furthermore, the systems have uniform signal input powers ranging from -10 to +10 dBm with steps of 1 dBm. For each case, the number of iterations is reported in the table below, where divergence and infinite oscillation are indicated with Div and Osc respectively:

**Table 3.1**: Number of iterations until convergence
for different signal and pump powers for CH=5 and CL=0.1

| $P_{signal,in}$(dBm) \ Adjustment | 1 | 0.9 | 0.8 | 0.7 | 0.6 | 0.5 | 0.4 | 0.3 | 0.2 | 0.1 |
|---|---|---|---|---|---|---|---|---|---|---|
| -10 | 55 | 187 | 155 | 430 | 181 | 840 | Div | Div | Div | Div |
| -9 | 58 | 179 | 150 | 298 | 171 | 1211 | Div | Div | Div | Div |
| -8 | 61 | 172 | 142 | 169 | 453 | 586 | Div | Div | Div | Div |
| -7 | 66 | 160 | 145 | 169 | 540 | 980 | Div | Div | Div | Div |
| -6 | 73 | 215 | 197 | 676 | 348 | 751 | Div | Div | Div | Div |
| -5 | 108 | 59 | 192 | 361 | 270 | 702 | 1631 | Osc | Div | Div |
| -4 | 228 | 48 | 186 | 155 | 280 | 360 | 1507 | Osc | Div | Div |
| -3 | 219 | 85 | 180 | 146 | 254 | 470 | 868 | Div | Div | Div |
| -2 | 89 | 143 | 175 | 226 | 370 | 423 | 1457 | Div | Div | Div |
| -1 | 64 | 203 | 169 | 189 | 159 | 304 | 421 | Div | Div | Div |
| 0 | 99 | 195 | 164 | 138 | 366 | 186 | 510 | Div | Div | Div |
| 1 | 123 | 190 | 159 | 131 | 221 | 291 | 1069 | Osc | Div | Div |
| 2 | 135 | 183 | 155 | 145 | 138 | 255 | 581 | Osc | Div | Div |
| 3 | 123 | 177 | 151 | 135 | 169 | 251 | 376 | 2329 | Div | Div |
| 4 | 201 | 173 | 147 | 147 | 238 | 201 | 403 | 1467 | Div | Div |
| 5 | 197 | 170 | 145 | 117 | 171 | 335 | 553 | 877 | Div | Div |
| 6 | 194 | 168 | 143 | 115 | 159 | 177 | 320 | Osc | Div | Div |
| 7 | 192 | 167 | 142 | 113 | 180 | 154 | 581 | 1047 | Div | Div |
| 8 | 191 | 167 | 143 | 116 | 185 | 270 | 368 | 777 | Div | Div |
| 9 | 193 | 167 | 143 | 112 | 184 | 177 | Osc | 693 | Div | Div |
| 10 | 233 | 159 | 145 | 116 | 186 | 170 | Osc | 445 | Div | Div |

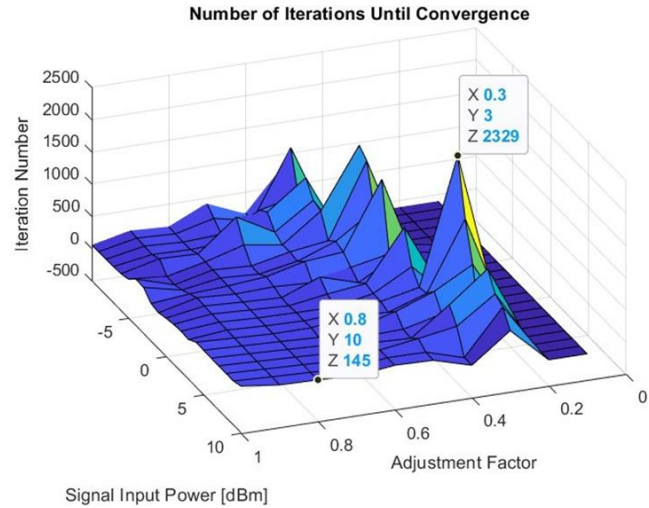

**Figure 3.1**: Iterations until convergence for the proposed method with CH=5, CL=0.1, $step_{dBm,signal}$ = 2, $step_{dBm,pump}$ = 0.5 and $factor_{signal}$ = 4 for various signal and pump power

Figure 3.1 is a surface plot visualizing the same data, in which a value of -100 is considered to signal divergence while a value of 0 is considered if oscillation happens. It can be seen that the proposed method is able to achieve convergence for most cases of *Adjustment* > 0.2. If a maximum total pump power of 3 Watts is considered, corresponding *to Adjustment*$_{min}$ = 0.4, and signal input power is bounded between -5 and 10 dBm, the proposed method converges in all but 2 cases; namely the ($P_{signal,in}$, *Adjustment*) couples (9, 0.4) and (10, 0.4) which can be considered extreme cases.

Furthermore, the average number of iterations taken by the method to achieve convergence in the filtered case mentioned is 263. For reference, 263 iterations are computed in less than two seconds using an Intel Core i7-1255U processor (1.7 GHz).

To search for better CH and CL parameters, the above test is repeated six times for values of CH in [1 3 5] and CL in [0.1 0.05]. Six matrices, similar to the one above, are obtained. The average number of iterations for convergence in each case, along with the number of converged cases, is reported in the Table 3.2 and 3.3:

Table 3.2: Average iterations for different CH and CL settings

| CH | CL = 0.1 | CL = 0.05 |
|---|---|---|
| 1 | 395.6733 | 367.7152 |
| 3 | 313.6776 | 352.7417 |
| 5 | 321.5197 | 395.6340 |

Table 3.3: Number of converged cases for different CH and CL settings

| CH | CL = 0.1 | CL = 0.05 |
|---|---|---|
| 1 | 150 | 151 |
| 3 | 152 | 151 |
| 5 | 152 | 153 |

It can be understood that the values of the correction factors CH and CL do not have a big effect on convergence as it is achieved for almost the same number of cases across all values considered. Moreover, the pair CH = 3, CL = 0.1 seem to achieve a slight gain in efficiency (through average iteration number) with respect to the case CH = 5, CL = 0.1. Consequently, it is beneficial to switch to CH = 3 in the method parameters, which will be applied in the following tests.

This study shows how the method is able to provide a solution for the Raman coupled differential equations in most cases of a C+L system with uniform input power. It also shows that the method does so in a relatively low average number of iterations, completing the initial objective.

## 3.2 Comparison With Conventional Method

In this section, a comparison between the developed method and the conventional method based on MATLAB's bvp4c function is presented. The focus will be on the speed of each method in different UWB systems, such as the C+L, C+L+S and C+L+S+E systems. the relationship between increasing bandwidth and the efficiency difference between the two methods will be explored. Moreover, convergence will be compared, highlighting cases where the proposed method successfully converges in situations where the conventional method fails.

### 3.2.1 Speed

**(1) C+L systems with uniform signal input power**

Table 3.1 showed that the proposed method takes around 260 iterations on average to converge for a C+L system. To see how much this method improves over the conventional one a study similar to the one reported in Table 3.4 is done, with reduced cases because the conventional method is computationally expensive. The Adjustment values considered are [1 0.7 0.4] while those of $P_{signal,in}$ are [-5 0 5 10]. The conventional bvp4c method is ran three times for each case and the elapsed time is averaged to reduce the impact of measurement noise. The same is done with the proposed method for comparison. The results are presented in the following tables.

Table 3.4: Average time elapsed by the proposed method for varying signal and pump power

| $P_{signal,in}$ (dBm) | Adjustment = 1 | Adjustment = 0.7 | Adjustment = 0.4 |
|---|---|---|---|
| -5 | 0.4388 | 2.7875 | 4.9895 |
| 0 | 0.7254 | 1.1965 | 2.9472 |
| 5 | 1.5989 | 0.5950 | 1.9007 |
| 10 | 1.5121 | 0.8363 | osc |

**Table 3.5**: Average time elapsed by the conventional bvp4c method for varying signal and pump power

| Psignal,in(dBm) | Adjustment = 1 | Adjustment = 0.7 | Adjustment = 0.4 |
|---|---|---|---|
| -5 | 11.8595 | 13.4503 | 17.2416 |
| 0 | 18.2212 | 26.9112 | 23.5033 |
| 5 | 13.4306 | 15.9686 | 24.7477 |
| 10 | 22.5177 | 16.5569 | 25.8477 |

**Table 3.6**: Error Matrix for Different Adjustment Factors and Signal Powers

| Psignal,in(dBm) | Adjustment = 1 | Adjustment = 0.7 | Adjustment = 0.4 |
|---|---|---|---|
| -5 | 11.8595 | 13.4503 | 17.2416 |
| 0 | 18.2212 | 26.9112 | 23.5033 |
| 5 | 13.4306 | 15.9686 | 24.7477 |
| 10 | 22.5177 | 16.5569 | 25.8477 |

The error between the 2 methods is reported in Table 3.6. The amount of gain in time between the 2 methods is seen to be substantial. To quantify it, the time gain is defined as the ratio of the time elapsed, shown in Table 3.7.

**Table 3.7**: Time gain between the conventional and proposed methods for varying signal and pump power

| Psignal,in (dBm) | Adjustment = 1 | Adjustment = 0.7 | Adjustment = 0.4 |
|---|---|---|---|
| -5 | 27.03 | 4.83 | 3.46 |
| 0 | 25.12 | 22.49 | 7.97 |
| 5 | 8.40 | 26.84 | 13.02 |
| 10 | 14.89 | 19.80 | NaN (osc) |

The gain is substantial, it seems to get lower for increasing pump power (decreasing Adjustment factors) and for lower signal input power. This can be explained by the fact that the proposed method would need to scale down the pumps in cases of high pump power, loosing efficiency.

The values of table 3.7 are averaged for all the cases to obtain the average time gain achieved by the proposed method over all C+L cases considered. That value is 15.80.

### (2) C+L systems with non-uniform signal input power

The developed method's performance is now compared to the conventional one in the case of more realistic C+L systems where the signal power is not uniform but varying according to the following figure:

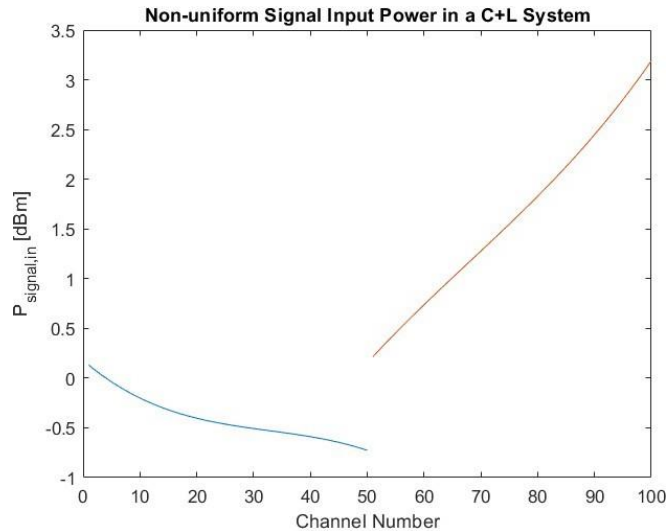

**Figure 3.2**: Non-uniform signal input power C+L system

The varying signal input power illustrated above is applied to the same C+L system used before. The pumps used are the same as the previous sections with *Adjustment* = 1. The output of both methods is shown below, along with the time elapsed averaged over five tries and the error:

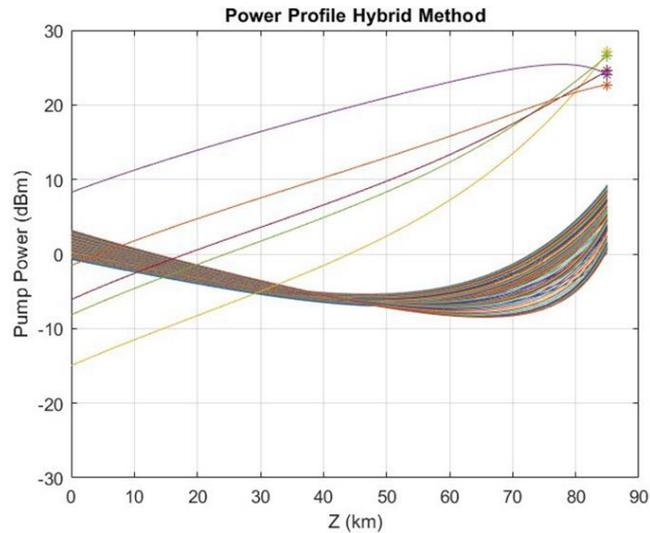

**Figure 3.3**: proposed method output for non-uniform signal input power C+L system

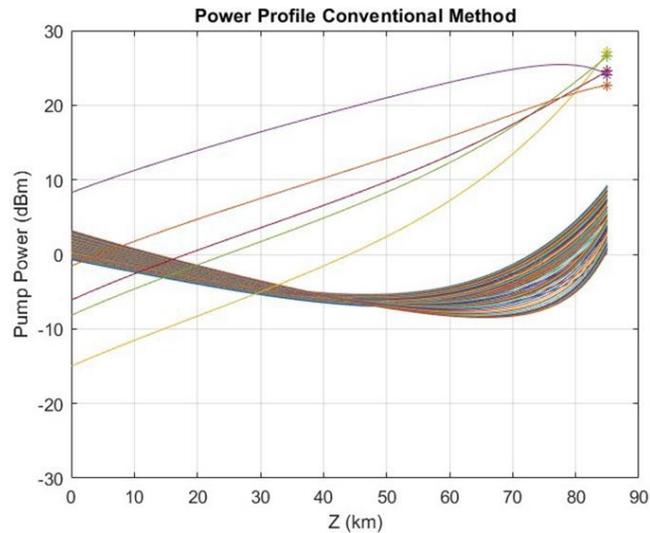

**Figure 3.4**: Conventional method output for non-uniform signal input power C+L system

The two figures look similar, in fact the error between the two is 0.0352 dBm, an accepted one. More importantly, the average time for convergence of the proposed method is 0.910 seconds while the conventional one takes 23.307 seconds to converge, representing an average time gain of approximately 25.

To further test this time gain, the experiment is repeated for different scenarios of non-uniform signal input power. The input powers shown in Fig. 3.2 are scaled by a factor k where it takes values in [1 2 3] while the Adjustment factor is varied between the values [1 0.7 0.4]. The average times taken by both methods after 3 runs of the 9 scenarios are reported below.

**Table 3.8**: Average timing of the proposed method for non-uniformly powered C+L systems

| k Factor | Adjustment = 1 | Adjustment = 0.7 | Adjustment = 0.4 |
|---|---|---|---|
| 1 | 0.5873 | 1.3039 | 4.9541 |
| 2 | 0.7754 | 1.0539 | 2.0868 |
| 3 | 1.0778 | 1.3084 | 2.3356 |

**Table 3.9**: Average timing of the conventional method for non-uniformly powered C+L systems

| k Factor | Adjustment = 1 | Adjustment = 0.7 | Adjustment = 0.4 |
|---|---|---|---|
| 1 | 30.1953 | 46.2053 | 48.8202 |
| 2 | 35.2815 | 40.0672 | 42.4700 |
| 3 | 25.4631 | 54.1620 | 39.7435 |

**Table 3.10:** Error Matrix for non-uniformly powered C+L systems

| k Factor | Adjustment = 1 | Adjustment = 0.7 | Adjustment = 0.4 |
|---|---|---|---|
| 1 | 0.0611 | 0.0304 | -27.9443 + 13.6438i |
| 2 | 0.0620 | 0.0364 | -27.3306 + 13.6438i |
| 3 | 0.0591 | 0.0395 | -26.3177 + 13.6438i |

The error between the 2 methods is reported below, where complex values indicate the non-convergence of the conventional "bvp4c" method. This will be further studied later. Over all the scenarios, the proposed method is able to outperform the conventional one in non-uniform signal input power C+L systems with an overall average time gain of 31.40.

### (3) C+L+S systems

At this stage, the performance of the method is compared for wider UWB systems, namely ones with the added S band. This band takes the range of frequency between 196 and 201 THz. Along with the C+L channels considered before, 38 channels are added in the range of the S band with a channel spacing of 125 GHz and a symbol rate of 100 GBaud. The signal channels are powered in input in a realistic non-uniform fashion as illustrated by the figure below.

Moreover, the pumps are similar to the ones used before with the same power scalable by the Adjustment factor, however their frequencies are shifted by 5 THz to make room for the S band. The pump frequencies thus look like this: [215.56, 213.87, 211.72, 209.51, 205.55] THz.

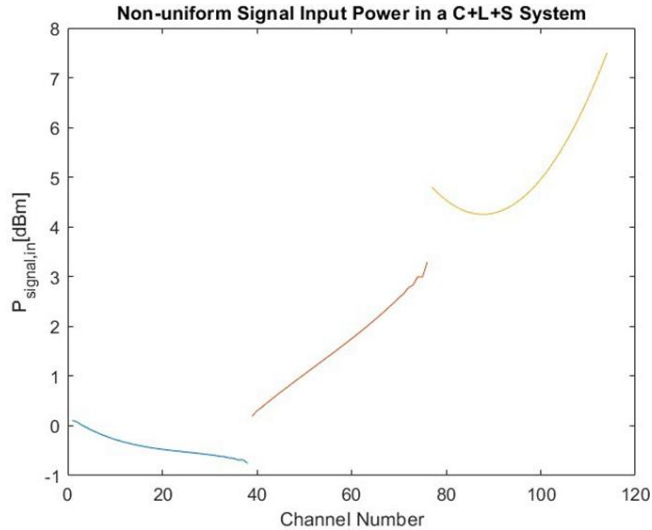

**Figure 3.5**: Non-uniform signal input power C+L+S system

This system's power profile throughout the span of the fibre is solved four times with different boundary conditions. Namely by varying the Adjustment factor between the values [1 0.5] and the k factor, introduced in

2.5, between [1 2]. The time elapsed by both methods, averaged over 3 runs is reported below, along with the error. Consequently, the average time gain achieved over all runs of C+L+S systems, is 38.009.

Table 3.11: Timing of the proposed method for non-uniformly powered C+L+S systems

| k Factor | Adjustment = 1 | Adjustment = 0.5 |
|---|---|---|
| 1 | 0.6933 | 1.3664 |
| 2 | 0.9243 | 1.5358 |

Table 3.12: Timing of the conventional method for non-uniformly powered C+L+S systems

| k Factor | Adjustment = 1 | Adjustment = 0.5 |
|---|---|---|
| 1 | 35.9251 | 52.1826 |
| 2 | 27.8826 | 48.9343 |

Table 3.13: Error Matrix for non-uniformly powered C+L+S systems

| k Factor | Adjustment = 1 | Adjustment = 0.5 |
|---|---|---|
| 1 | 0.0603 | 0.4079 |
| 2 | 0.0482 | 0.0323 |

To conclude, the method for power profile computation developed throughout this paper achieves substantial time gain over the bvp4c method of MATLAB in all possible cases. A summary of all the time gains is reported below:

Table 3.15: Average Time Gain in Different Scenarios

| Scenario | Average Time Gain |
|---|---|
| C+L Uniform | 15.80 |
| C+L Non-Uniform | 31.40 |
| C+L+S Non-Uniform | 38.01 |

It is thus concluded that the efficiency of the proposed method increases for increasing signal input power complexity and increasing bands. This is hypothesized to be due to the fact that with increasing bands, the only thing different for the developed method is the size of the matrices in Eq. (8). Whereas other methods that try to solve the differential equations directly, would have an increasing number of coupled equations which would be increasingly less efficient to solve. This makes the proposed method an even more important step forward towards the efficient power profile computation in the context of ultra-wide band systems.

Finally, an explanation of the reason why the proposed method seems to be so much faster than the bvp4c method is given. For solving the coupled Raman differential equations, the conventional method relies on MATLAB's bvp4c function, which addresses the boundary value problem by solving two initial value problems, one forward and one backward from opposite ends of the fiber. The method iterates by adjusting initial guesses until the solutions from both directions converge. According to the solver's output, the solution is reached after 148068 calls to the ODE function and 494 calls to the boundary condition function.

Each iteration of the conventional method involves complex numerical calculations, which accumulate over a large number of iterations. In contrast, the proposed method developed in this work achieves the same results through efficient matrix multiplications and requires only a few hundred iterations on average. This significant efficiency difference arises from the proposed method's ability to reduce the computational complexity per iteration, and number of iterations, thus offering substantial time savings while maintaining accuracy.

### 3.2.2 Convergence

The efficiency study has already shown that both methods converge for most of the systems presented. The error Tables 3.6, 3.10 and 3.13 showed the maximum error between the power profiles computed by each of the methods to be negligible in most cases indicating that both are accurate enough. However, in some cases the error is a complex number. This indicates that one of the methods did not converge.

Consequently, to compare the two methods in terms convergence, an example is taken where the error is complex. A case from table 3.10 is taken where k = 2 and *Adjustment* = 0.4. The error was shown to be complex with high magnitude. In fact, the power profile from the conventional method is shown in Fig. 3.6, while the solution provided by the proposed method is reported in Fig. 3.7.

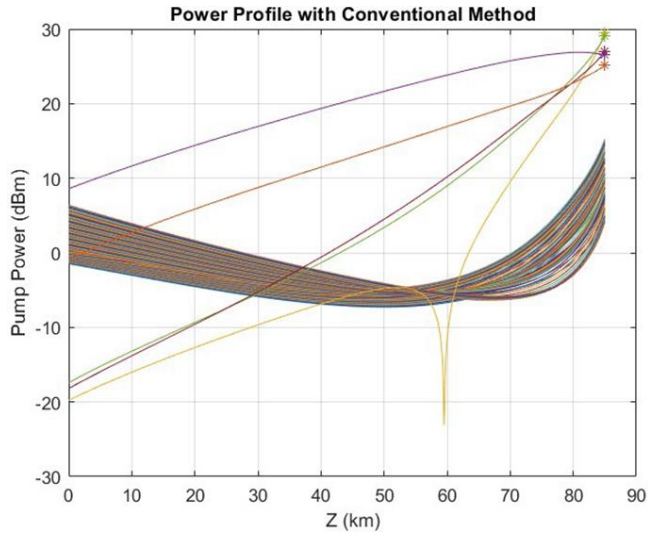

**Figure 3.6:** Non-uniform signal input power C+L system with conventional method

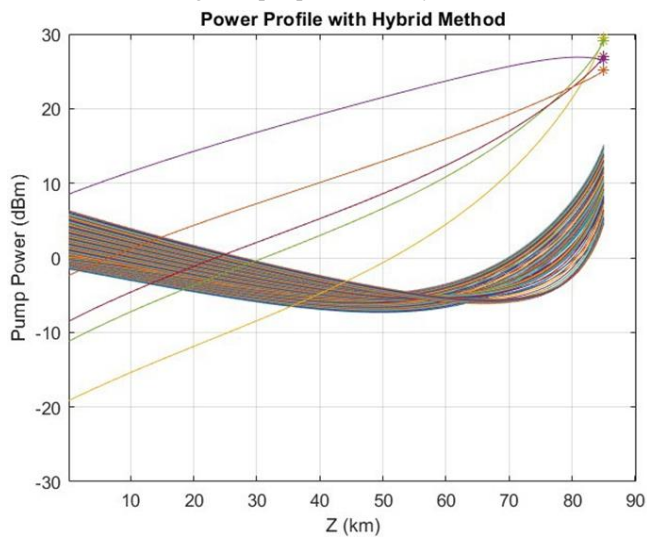

**Figure 3.7**: Non-uniform signal input power C+L system with proposed method

It is immediately seen that the output of the conventional method is erroneous, because of the dip in the power of one of the pumps. In fact, this pump has a complex profile inside the matrix P. This in turn affects the signals and the whole reliability of this method. On the other hand, the mew method provided a real solution.

As seen in the different error tables, this scenario happens multiple times. In all of which the proposed method is able to provide real solutions while the complex values arise in bvp4c. It is concluded that the proposed method not only outperforms bvp4c in efficiency but also converges in a wider range of systems.

Through extensive testing, it has been demonstrated that the method developed in this work is significantly more efficient than the conventional approach previously used. On average, it achieves a time gain factor of around 30, depending on the specific scenario. Additionally, the proposed method has shown robust convergence across a wide variety of cases, even in situations where the conventional method fails. This makes the developed approach superior in all critical aspects, particularly in terms of efficiency and reliability.

## 4. Conclusion

In this paper, a more efficient, reliable, and adaptive method for power profile calculation in backward Raman amplified system was developed, leading to a significant speedup of the EGN closed-form model, paving the way for real-time application in commercial systems.

The algorithm demonstrated robustness by achieving convergence in all practical scenarios, even in cases where conventional methods failed. Moreover, it proved to be significantly more efficient with no loss in accuracy, delivering an average time factor improvement of 30 compared to the existing method. Furthermore, it was shown that the efficiency of the developed method increases with the complexity and the bandwidth of the system. This advancement enabled optimizations of multi-band systems to be performed in a fraction of the time required by conventional approaches. Such improvements will make it possible to conduct complex optimizations on increasingly larger UWB systems much faster, accelerating research in this field and driving the commercial viability of these technologies.